# Applying Device-to-Device Communication to Enhance IoT Services

Ji Lianghai, *Student Member, IEEE*, Bin Han, *Member, IEEE*, Man Liu and Hans D. Schotten, *Member, IEEE*

*Abstract*—Massive Machine Type Communication (mMTC) to serve billions of IoT devices is considered to open a potential new market for the next generation cellular network. Legacy cellular networks cannot meet the requirements of emerging mMTC applications, since they were designed for human-driven services. In order to provide supports for mMTC services, current research and standardization work focus on the improvement and adaptation of legacy networks. However, these solutions face challenges to enhance the service availability and improve the battery life of mMTC devices simultaneously. In this article, we propose to exploit a network controlled sidelink communication scheme to enable cellular network with better support for mMTC services. Moreover, a context-aware algorithm is applied to ensure the efficiency of the proposed scheme and multiple context information of devices are taken into account. Correspondingly, signaling schemes are also designed and illustrated in this work to facilitate the proposed technology. The signaling schemes enable the network to collect required context information with light signaling effort and thus network can derive a smart configuration for both the sidelink and cellular link. In order to demonstrate the improvements brought by our scheme, a system-level simulator is implemented and numerical results show that our scheme can simultaneously enhance both the service availability and battery life of sensors.

## I. INTRODUCTION

MASSIVE Machine Type Communications (mMTC) [1], [2] is widely considered as an important service to be offered by the upcoming fifth generation (5G) cellular networks. mMTC refers to a typical Internet-of-Things (IoT) scenario, where a large amount of static sensors are deployed and report sporadically to an application server in the cloud (e.g., to enable environment monitoring and object condition tracking). While opening a new potential market, mMTC also poses different requirements on network. Since legacy cellular networks were designed for services with high data rate and low latency, they experience technical challenges to meet the requirements of mMTC services (e.g., low device cost, long device battery life and high service availability). In order to obtain a deep market penetration in exploiting 5G to support mMTC services, the Third Generation Partnership Project (3GPP) has conducted studies to adapt and evolve legacy networks. For instance, to reduce device complexity, a new type of user equipment (UE) is introduced as the category 0 in [3], which has a reduced peak data rate, a single antenna design and reduced bandwidth. Further cost reduction can be gained by reducing the maximal transmission power [2] to simplify the integration of power amplifier (PA). However, it reduces the network coverage in uplink as a trade-off. Besides, since some mMTC devices are deployed in deep indoor scenarios, an extra penetration loss up to 20 dB can be foreseen [4]. In order to maintain the uplink coverage, both narrow band transmission and massive transmission time interval (TTI) bundling [4] can help in serving deep coverage area, but lead to huge time resource usage on the system level and battery drain on the device level. It is, therefore, important to have a mechanism that is more efficient in power and resource to support wide deployment scenarios.

As one of the critical technical enablers for 5G networks, device-to-device (D2D) communication [5] provides an efficient alternative to cope with the requirements of mMTC services. The original motivation of exploiting D2D communication was to enable local information exchange for high reliability and low latency [5]–[7]. In 3GPP, D2D communication has been standardized to enable discovery and communication between two D2D devices [8]. Specifically, UE-to-network relaying is introduced in LTE release 13 to extend network coverage in public safety scenario, to assist automatic driving and achieve high transport efficiency. In that scheme, base station (BS) can control the selection of relay UEs by setting requirements on radio quality that a relay UE should fulfill. Moreover, the ranking of different UE-to-network relays is performed locally at remote UE and the remote UE may select the relay with the best radio quality to itself. Since the features above are designed to provide public safety, they are not optimized for the mMTC. Recently, this technology has been studied to enhance the mMTC services. In [9] [10], it is proposed that one cellphone can act as a relay with D2D links to other sensors. Thus, packets generated by those sensors can be transmitted to BS through the relay. In 3GPP, the possibility to exploit D2D communication for mMTC applications is also

L. Ji, B. Han, M. Liu and H. D. Schotten are with Institute for Wireless Communication and Navigation, Department of Electrical and Computer Engineering, University of Kaiserslautern, 67663 Kaiserslautern, Germany.
Emails: {ji,binhan,manliu,schotten}@eit.uni-kl.de

Figure 1: Our proposed approach allows the sensors with better propagation channel and enough battery level to relay uplink packets from other sensors.

studied in [11] where one cellphone acts as a relay for a group of mMTC sensors. Since the D2D pairing procedure is performed in a distributed manner without help from BS, it brings a loss in global awareness. Additionally, very few cellphones will appear in deep indoor or rural areas, especially at night. Thus, the proposed approaches have limited applicability.

In this work, we refer the sensors in cell border as remote sensors, since they experience high pathloss for cellular links. We inspect on how to improve the availability of the remote mMTC sensors and meanwhile also attempt to have more sensors fulfilling the battery life requirement. A scheme is proposed where sidelink communication can be performed among mMTC sensors. In this way, remote sensors have the opportunity to efficiently set up sidelinks with relay sensors. It worths to note that, in 3GPP, the term "sidelink" refers to a direct link between two devices over air interface, while the term "D2D communication" normally refers to a general system structure to support the direct link. This article provides some insights for the future standardization work in 3GPP to apply D2D communication in mMTC. Our solution differs from the work of LTE release 13 [8] at following points:

- Our focus is the smart transmission mode (TM) configuration of relay and remote UEs in mMTC scenario. So far, the corresponding technology is designed in 3GPP for public safety scenario.
- In our scheme, D2D groups are pre-selected by BS, while the current approach in 3GPP is that the remote UE locally performs the ranking of the potential UE-to-network relays.
- mMTC-related context information (e.g., sensor location, battery level) are collected with reasonable signaling overload to improve service availability and power consumption of sensors.

To describe the proposed scheme, our article is organized as following. At first, the system model for the proposed scheme is introduced and then a context-aware clustering scheme is provided to guarantee the sidelink transmission efficiency. Following that, a smart TM selection procedure is also given, in order to configure each sensor properly. Afterward, signaling schemes to support the proposed technology are illustrated. Additionally, to compare the performance of the new scheme with LTE network, a system level simulator is implemented and numerical results are shown.

## II. SYSTEM MODEL

Many mMTC services request sensors to sporadically report their status with a short payload in uplink. Downlink transmission is used to transmit system and control information. Due to a higher available transmission power at BS, mMTC services have better coverage in downlink than in uplink. Therefore, in this work, we ignore the downlink performance but only inspect the uplink performance of mMTC services. Besides, we consider mMTC sensors as static and able to perform sidelink communication with other sensors directly. Figure 1 shows the proposed scheme by applying sidelink to enhance mMTC services. Sensor #2 experiences a better propagation channel to BS than sensors #4 and #5, due to a smaller distance to the BS. If the battery level of sensor #2 is high enough, it can be configured by the BS to set up sidelink communication with sensors #4 and #5, and thus relay their data packets to the BS. In order to improve the performance, context information should be collected and exploited by the network to set up sidelink communication. For instance, the sensors in virtual cluster #m should not have direct sidelinks with the ones in virtual cluster #n, since they are geometrically separated. As the figure illustrates, three TMs exist for mMTC sensors.

- Cellular TM, where sensors upload their reports to BS via cellular links.
- Relay TM, where sensors receive reports from other sensors and then transmit both the received packets and their own reports to BS.
- Sidelink TM, where sensors transmit reports to corresponding relay sensors.

## III. DEVICE CLUSTERING AND TM SELECTION

As aforementioned, to achieve a high efficiency, sidelink communication should only happen between sensors located closely to each other. An efficient approach here is to perform device clustering algorithm and allow only intra-group sidelink communication. Thus, the context-aware sidelink communication scheme comprises of two steps, as:

1) clustering of mMTC devices;
2) selection of TM.

Once a sensor is initially attached to the network or the network needs to update the TMs of the serving sensors, the above two steps should be carried out in BS.

### A. Clustering of mMTC devices

Since a sidelink is applied for two nearby devices, an efficient clustering algorithm can assist remote sensors to find proper relay sensors. The design and implementation of an efficient clustering should take account of useful context information, which refers to the information (e.g., location information, traffic type, battery life requirement, etc) that can help to improve system performance. For instance, remote sensors in rural area are located far away from the BS. In this sense, by analyzing the location information of sensors, a sensor located between the remote sensor and the BS can be recognized and considered as a relay. In contrast, in a dense urban scenario, sensors located deep indoor or in basements are seen as remote sensors due to the high penetration loss. Therefore, the relay sensors, in this case, may be the sensors located on higher building floors. In this work, our proposed scheme is inspected in a rural area scenario. Thus the clustering algorithm is adapted correspondingly. At first, BS converts the location of sensors to their reference angles. And then the $K$-means clustering algorithm w.r.t. the reference angles of sensors is performed in BS. The basic steps of the $K$-means algorithm are listed below:

1) Initially, BS selects $K$ sensors with reference angles separated from each other as far as possible, these sensors are considered as centroids of $K$ clusters.
2) Take another sensor and associates it to the cluster, whose centroid is the closest to the selected sensor w.r.t. reference angle.
3) Calculate the mean reference angle of the updated cluster, and select the sensor in this cluster nearest to the mean value as the new centroid.
4) Repeat steps 2) and 3) until every sensor is associated with a cluster.

Since sensors are assumed to be static in our considered mMTC scenario, BS can acquire their location at their initial attachments to the network, as stated in the next section.

*B. TM selection*

Once a cluster is formed, BS configures every sensor in this cluster to a proper TM. Context information is helpful to optimize the TM of each device. For instance, a higher pathloss of a cellular link introduces a higher power consumption per uplink transmission. Thus, the battery life requirement might not be met for this sensor and BS tries to assign a relay to improve its battery usage. Moreover, some sensors even cannot reach the BS in uplink due to their extreme high pathloss values, therefore they are also assigned to a relay sensor which can contribute to improving the service availability. From the efficiency perspective, not all sensors with good cellular links are suitable to act as relays, since the sidelink radio quality also plays a critical role in power consumption. Besides, since each relay multicasts a sidelink discovery message to neighbor nodes during the sidelink setup or update procedure, the number of relay sensors should not be too high. Otherwise, more time and frequency resource are required for the sidelink discovery process. In this work, with the help of context information, the transmission mode selection (TMS) of sensors in each cluster is performed as following:

1) Sensors out of the uplink coverage and sensors with battery life lower than requirement are configured to the sidelink TM.
2) Sensors with good cellular links and enough battery capacity are considered as relay sensor candidates.
3) A relay sensor candidate with distance to the BS larger than a given threshold is selected as the relay sensor.
4) The configuration message is sent to both relay and remote sensors to trigger the sidelink discovery procedure.
5) In sidelink discovery procedure, a remote sensor is configured to sidelink TM, if its sidelink channel pathloss is below a threshold. Otherwise, it remains in cellular TM.
6) Other sensors not involved in sidelink communication are configured to cellular TM.

To be noticed, in rural areas, the remote sensors mentioned in step 1) are located far away from the BS. Thus, relay sensor candidates with distance to BS larger than a given radius, as stated in step 3), can ensure them to be in the proximity of the remote sensors. Taking its power consumption and signaling effort into account, the period of sidelink update procedure should be larger than the minimal uplink transmission period of sensors. Otherwise, multiple sidelink update procedures may take place between two consecutive reports of a sensor, but only the last sidelink update procedure makes sense. On the other hand, increasing the update period reduces the network flexibility to respond to condition changes. For instance, since a relay sensor forwards the packets of a group of sensors, it experiences a high battery power drain. Therefore, the network needs to update the sidelinks in a timely manner, to prevent from the case where a relay sensor powers out and the remote sensors will lose connections until the next sidelink update procedure takes place.

IV. Radio link enabler to Support the Corresponding Scheme

To support the proposed scheme, corresponding signaling procedures are introduced in this section. Three important procedures, including initial attachment of sensors, the update of TM and uplink transmission exploiting sidelink communication, are provided with details.

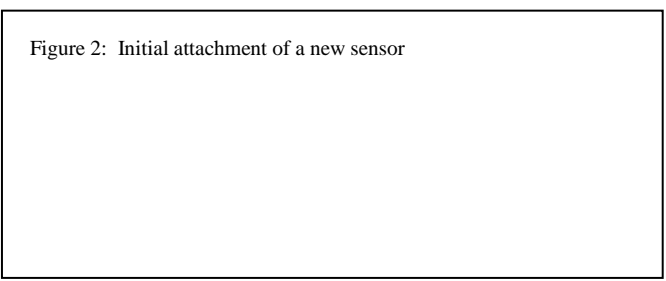

Figure 2: Initial attachment of a new sensor

*A. Initial attachment of a sensor*

Once a sensor is deployed and powered on, an initialization procedure shown in Fig. 2 is required. The signaling diagram is illustrated with more details in the following:

1) The sensor receives the sidelink system information blocks (S-SIBs), then reports its channel state information (CSI), location and battery information to the BS. Since the deployment of the new sensor is performed manually by technicians, such information can be submitted by equipment of technicians though the device might be in the uplink outage. Other context information such as traffic type can be requested by the

   BS from the corresponding application server. Afterward, BS will perform the sidelink clustering and TMS algorithm, taking account of the collected context information.
2) In case the sensor is configured to either relay or sidelink TM, the BS transmits the ID of its virtual cluster and other control information for sidelink discovery and communication to this sensor.
3) If the new sensor is configured to sidelink TM with sensor #m as the corresponding relay, sensor #m sends a sidelink discovery announcement message containing the ID of the new sensor. In another case, if the new sensor is configured to relay TM for sensor #m, it transmits the sidelink discovery message to sensor #m, where the ID of sidelink counterpart is also carried. The time and frequency resources used to transmit the discovery message are indicated in step 2). In this step, reference symbols to estimate sidelink channel will also be transmitted.
4) Upon receiving the sidelink discovery message, the receiver sensor determines whether the discovery request should be accepted, based on the estimated sidelink reference signal received power (S-RSRP). An acknowledgment (ACK) / non-acknowledgment (NACK) message is transmitted back.
5) If the discovery procedure is acknowledged, a security association between the two sidelink ends can be established by exchanging security-related messages.
6) The discovery decision is further forwarded to the BS by the relay sensor, thanks to its good cellular channel condition. In case a NACK message is received, the BS will configure the remote sensor to a cellular TM and avoid setting up the same sidelink ends in future.

In the procedure above, the S-SIBs provide the less frequently changed configuration information for sidelink (e.g., resource pools used for sidelink discovery and communication). Thus, the relay sensor transmits the discovery announcement message on the configured resource so that the monitoring sensors can receive and process the messages from the same resource. Moreover, since both the sidelink ends are already synchronized to the same access node in downlink and they are statically located within the proximity of each other, no synchronization procedure between them is needed. Last but not least, certain context information (e.g., security information) should be stored at both the sidelink ends once the sidelink connection is established. In this way, a fast sidelink connection resumption can be obtained with less signaling effort, in order to efficiently support the sporadic traffic of mMTC services.

*B. TM update procedure*

In case some conditions are no more fulfilled for sidelink communication (e.g., the battery level of a relay sensor decreases severely and is not sufficient to support the relay functionality anymore), TMs of the sensors in that cluster need to be updated. The corresponding signaling diagram is provided in Fig. 3 where two sensors are to be updated with the sidelink TM. This scheme can be easily extended to the case where more sensors are assigned with sidelink TM. The signaling diagram is illustrated with more details in the following:
1) BS performs the context-aware TMS algorithm.
2) BS pages the corresponding sensors and configures them with their new TMs. Besides, other dedicated control information is also transmitted (e.g., IDs of the sensors involved in sidelink, resource for sending sidelink discovery message). In Fig. 3, sensors #m and #n are configured to sidelink TM and another sensor is configured to relay TM.
3) The relay sensor sends a discovery announcement message to the sensors in sidelink TM, with the IDs of itself and the conveyed target sensors. Reference signals are embedded in this message.
4) Based on the S-RSRP, receivers of the discovery message (i.e. sensors #m and #n) determine whether the sidelink request is accepted. The decision will be fed back to the relay sensor.
5) If the request is accepted, a security association is established.
6) The results of TM update are further transmitted from the relay sensor to the serving BS. If a sensor fails in establishing sidelink with the relay sensor, BS configures it to cellular TM and avoids to pair these two sensors for sidelink communication in future.

Figure 3: Procedure for TM update

As the relay sensor transmits a discovery announcement to a group of remote sensors in step 3), each sensor in step 4) can respond by picking up a resource from the resource pool, which is indicated in the configuration information in step 2). Moreover, if a sidelink is successfully established in step 5), the sensors in sidelink TM should be aware of the discontinuous reception (DRX) cycle of the relay sensor, in order to derive the time instance that the relay sensor wakes up to receive the packets.

*C. Reports transmitted by sidelink communication*

In case certain sensors in sidelink TM are paged by network or have data in their buffer to transmit, the established sidelinks are exploited to relay the data in the uplink. Figure 4 shows this transmission procedure and it is further illustrated in the following:

1) Sensors involved in sidelink communication receive the S-SIBs.
2) In the mobile terminated case, one or multiple sensors in sidelink TM will be paged to trigger their reporting procedure by the BS. In the paging message, dedicated resource for the sidelink communication can be assigned. In Fig. 4, sensor #m, sensor #n and the corresponding relay sensor are paged by network to transmit their data reports.
3) In the mobile originated case, sensors in sidelink TM (i.e., sensor #m and sensor #n in Fig. 4) try to transmit their data packets through sidelink. There are two options to obtain resource for the sidelink transmission. One is that multiple remote devices can randomly access to the relay sensor by sending different preambles and the relay sensor replies with the assigned resource. Another option refers to a semi-persistent resource allocation scheme where the time and frequency resource for the sidelink transmission of each sensor is pre-allocated.
4) After successfully receiving packets from remote sensors (i.e. sensors #m and #n in Fig. 4), relay sensor replies with ACK messages back. Otherwise, a NACK message is transmitted back and triggers the re-transmission procedure.
5) Relay sensor further forwards the successfully received packets to the serving BS. This process can be performed as a normal cellular uplink transmission where a control plane (CP) connection needs to be established. Further, network can configure the relay sensor to compress its own uplink packet together with the packets from remote sensors and send them together to the BS. In this case, less power is consumed at relay sensor to transmit both its own packet and the packets from remote sensors, since the relay sensor needs to wake up and perform the CP connection establishment procedure only once.
6) Upon successfully receiving packets from the relay sensor, BS sends an ACK message back. Otherwise, a re-transmission procedure is triggered by a NACK message.

In order to update the context information at BS, real time sensor information such as battery level should be transmitted in the data reports from both the relay and remote sensors.

## V. EVALUATION AND NUMERICAL RESULTS

Figure 4: Uplink report procedure

TABLE I: Device power consumption parameters

(a) CDF of served days of all deployed mMTC devices    (b) CDF of served days of mMTC devices from two different sets (in the coverage or outage of LTE)
Figure 5: System performance of mMTC devices w.r.t. their served days in uplink

For evaluation, a system level simulator is used where the service availability and battery life of mMTC sensors are taken as key performance indicators. In the simulation, one single BS is deployed in rural area with a cell radius of 2500 meters, as defined in [12]. One hundred thousand sensors are uniformly distributed inside the cell at a height of 0.5 meter and each sensor has a maximal transmission power of 20 dBm, as proposed for LTE-M [2]. 900 MHz is assumed to be the carrier frequency for the air interface. Moreover, since different subbands are used for cellular link and sidelink, there is no mutual interference between them. The model proposed in [13] is implemented for the sidelink channel. Regarding the traffic model, sensors are required to periodically transmit their reports with a payload of 1000 bits, at a frequency of one report per 150 seconds. BS will consider a sensor as a relay candidate, only if this sensor has a distance to the BS larger than 1500 meters and meanwhile also a battery life which is 20 percent longer than the service requirement. Moreover, both cellular link and sidelink apply open power control scheme to achieve target signal-to-noise ratio (SNR) (3 dB for cellular link and 10 dB for sidelink). To evaluate battery life performance of mMTC sensors, parameters related to power consumption are given in Table I. Note that 10 ms is assumed to be the duration for CP connection establishment procedure. This is the value targeted by the 5G network while the legacy 4G network requires approximately 50 ms to set the CP connection. Other parameters (e.g., channel model for cellular link, noise figure, etc.) are aligned with the ITU-R document [14]. Last but not least, LTE technology is used to model radio link performance. In Fig. 5, cumulative distribution function (CDF) of served time of mMTC sensors is shown. As a baseline scheme, the performance of LTE release 12 is given where relay is not applied. Moreover, since LTE release 13 has already standardized the UE-to-network relaying for public safety scenario, its performance on the considered mMTC scenario is also provided. An SNR value of 6 dB in uplink should be fulfilled for being a relay node in both the LTE release 13 and our proposed context-aware D2D schemes. Additionally, both schemes schedule one hundred relays at each time, if there are enough relay candidates. In LTE network, if a radio link experiences a poor SNR (e.g., lower than -7 dB), no data transmission is possible on this link and the sensor is in the outage of network. In Fig. 5a, the performance of all deployed sensors is given. As it shows, more than six percent of sensors are in the outage of the BS in LTE release 12 and unable to transmit data to the BS in uplink. The discrete steps in the curve are caused by the fixed modulation and coding schemes (MCS) applied by LTE technology. Compared with the performance of LTE release 12, both LTE release 13 and our proposed context-aware D2D scheme improve the service availability so that all sensors can be served by either cellular link or sidelink. Additionally, 60 percent of the sensors have service life over 10 years in LTE release 12 and it can be improved to 78 percent by LTE release 13, while our proposed scheme raises this ratio to 95 percent. As aforementioned, relay sensors experience high power consumption and this is the reason that both the context-aware D2D and LTE release 13 schemes have fewer sensors with long serving days. For more detailed inspection on our proposed scheme, sensors are categorized into two different sets based on whether they are in the coverage of LTE network. In Fig. 5b, the system performances of two different sets are provided separately. Regarding the performance of LTE release 13, since no UE battery information is presented and controlled at the BS, relay UEs can run out of their battery very quickly. Therefore, it can be seen that the UEs in the coverage of the BS will have worse performance compared with the LTE release 12. On the other hand, due to the performance sacrifice of the relay UEs, the UEs out of BS coverage can be well supported in LTE release 13. Compared with the scheme in LTE release 13, the proposed context-aware D2D scheme contributes to a better control of both the relay (i.e., the dark blue curve) and remote UEs (i.e., the black curve). With respect to sensors in the outage of the BS of LTE release 12, they can be served by sidelink communication and 80 percent of them can even be served for more than 10 years. It should be noticed that, though these sensors are served by sidelink communication which consumes less power for data transmission than the cellular link, there are still 20 percent of users that cannot achieve the battery life requirement of 10 years. This is due to the fact that, since relay sensors experience high battery drain from forwarding the packets from remote UEs, they might lose the ability to be a relay after some time. Thus, after some time, there are not enough suitable relay sensors in the system and some remote UEs cannot be served by sidelink communication any more. Additionally, performance comparison of the sensors in the coverage of the BS is also shown in this figure, and we can see that the ratio of sensors meeting the battery life requirement is increased from 76 percent in LTE (i.e., the red curve) to 99 percent in the context-aware D2D scheme (i.e., the dark blue curve). Please note that, since the proposed context-aware D2D scheme targets at a battery life requirement of 10 years, the remote UEs will not be served by relay UEs any more, once this requirement has been fulfilled. This is the reason why a step from 1 percent to 24 percent occurs for the context-aware D2D scheme (i.e., the dark blue curve) at the point of 10 years, which shows that 23 percent of UEs are served exactly for 10 years by the context-aware D2D scheme.

## VI. Conclusion and Future Work

In this work, mMTC services are enhanced by a context-aware sidelink communication scheme. Three critical signaling procedures to support the proposed scheme are described, with respect to the initial attachment of sensors, update of TM and sidelink communication procedure. Meanwhile, the proposed signaling schemes enable the collection of required context information without a heavy signaling overhead. Moreover, the numerical results gained from a system level simulation show that our scheme improves both the service availability and the battery life of mMTC devices simultaneously.

## Acknowledgement

Part of this work has been performed in the framework of H2020 project METIS-II, which is funded by the European Union. The views expressed are those of the authors and do not necessarily represent the project. The consortium is not liable for any use that may be made of any of the information contained therein.

## References

[1] S.-Y. Lien, K.-C. Chen, and Y. Lin, "Toward Ubiquitous Massive Accesses in 3GPP Machine-to-Machine Communications," *IEEE Communications Magazine*, vol. 49, no. 4, 2011.
[2] "Study on Provision of Low-Cost Machine-Type Communications (MTC) User Equipments (UEs) Based on LTE (Release 12)," Third Generation Partnership Project (3GPP), Tech. Rep. TR 36.888, Jun. 2013.
[3] "User Equipment (UE) Radio Access Capabilities (Release 13)," Third Generation Partnership Project (3GPP), Tech. Rep. TR 36.306, Dec. 2016.
[4] "Cellular System Support for Ultra-Low Complexity and Low Throughput Internet of Things (CIoT) (Release 13)," Third Generation Partnership Project (3GPP), Tech. Rep. TR 45.820, Nov. 2015.
[5] K. Doppler, M. Rinne, C. Wijting, C. B. Ribeiro, and K. Hugl, "Device-to-Device Communication as an Underlay to LTE-Advanced Networks," *IEEE Communications Magazine*, vol. 47, no. 12, 2009.
[6] J. M. B. da Silva, G. Fodor, and T. F. Maciel, "Performance Analysis of Network-Assisted two-hop D2D Communications," in *Globecom Workshops (GC Wkshps)*, 2014. IEEE, 2014, pp. 1050–1056.
[7] A. Osseiran, J. F. Monserrat, and P. Marsch, *5G Mobile and Wireless Communications Technology*. Cambridge University Press, 2016.
[8] "Proximity-Based Services (ProSe) (Release 14)," Third Generation Partnership Project (3GPP), Tech. Rep. TS 23.303, Dec. 2016.
[9] N. K. Pratas and P. Popovski, "Underlay of Low-Rate Machine-Type D2D Links on Downlink Cellular Links," in *2014 IEEE International Conference on Communications Workshops (ICC)*, IEEE, 2014, pp. 423–428.
[10] ——, "Low-Rate Machine-Type Communication via Wireless Device-to-Device (D2D) Links," arXiv preprint arXiv:1305.6783, 2013.
[11] "New WI proposal: D2D Based MTC," Third Generation Partnership Project (3GPP), Tech. Rep. RP-151948, Dec. 2015.
[12] "Study on Scenarios and Requirements for Next Generation Access Technologies (Release 14)," Third Generation Partnership Project (3GPP), Tech. Rep. TR 38.913, Oct. 2016.
[13] "Discussion on UE-UE Channel Model for D2D Studies," Third Generation Partnership Project (3GPP), Tech. Rep. R1-130092, Jan. 2013.
[14] "Guidelines for Evaluation of Radio Interface Technologies for IMT-Advanced," ITU Radiocommunication Sector (ITU-R), Tech. Rep. M.2135, 2008.

**Ji Lianghai** (S'17) received his B.Sc degree from the Shandong University, China in 2010 and his M.Sc. degree from the University of Ulm, Germany in 2012. He is at this moment working for his Ph.D degree at the institute of Wireless Communication and Navigation, University of Kaiserslautern, Germany. He worked for European 5G flagship project METIS and some other 5G projects with industrial partners. He is currently representing University of Kaiserslautern in 5GPPP project METIS-2.

**Bin Han** (M'17) received his B.E. degree in 2009 from the Shanghai Jiao Tong University, China, and his M.Sc. degree in 2012 from the Technische Universität Darmstadt, Germany. In 2016 he was granted the Ph.D. (Dr.-Ing.) degree from the Kalsruhe Institute of Technology, Germany. He is currently a postdoctoral researcher at the Institute of Wireless Communication and Navigation, University of Kaiserslautern, Germany. His research interests are in the broad area of wireless communication systems and signal processing.

**Man Liu** received her Diplom degree from the University of Kaiserslautern, Germany in 2010. She is at this moment working for her Ph.D degree at the institute of Wireless Communication and Navigation, University of Kaiserslautern, Germany. Her research interests are on the application of 5G network for machine type communication.

**Hans D. Schotten** (S'93-M'97) received the Diplom and Ph.D. degrees in Electrical Engineering from the Aachen University of Technology RWTH, Germany in 1990 and 1997, respectively. Since August 2007, he has been full professor and head of the Institute of Wireless Communication and Navigation at the University of Kaiserslautern. Since 2012, he has additionally been Scientific Director at the German Research Center for Artificial Intelligence heading the "Intelligent Networks" department.

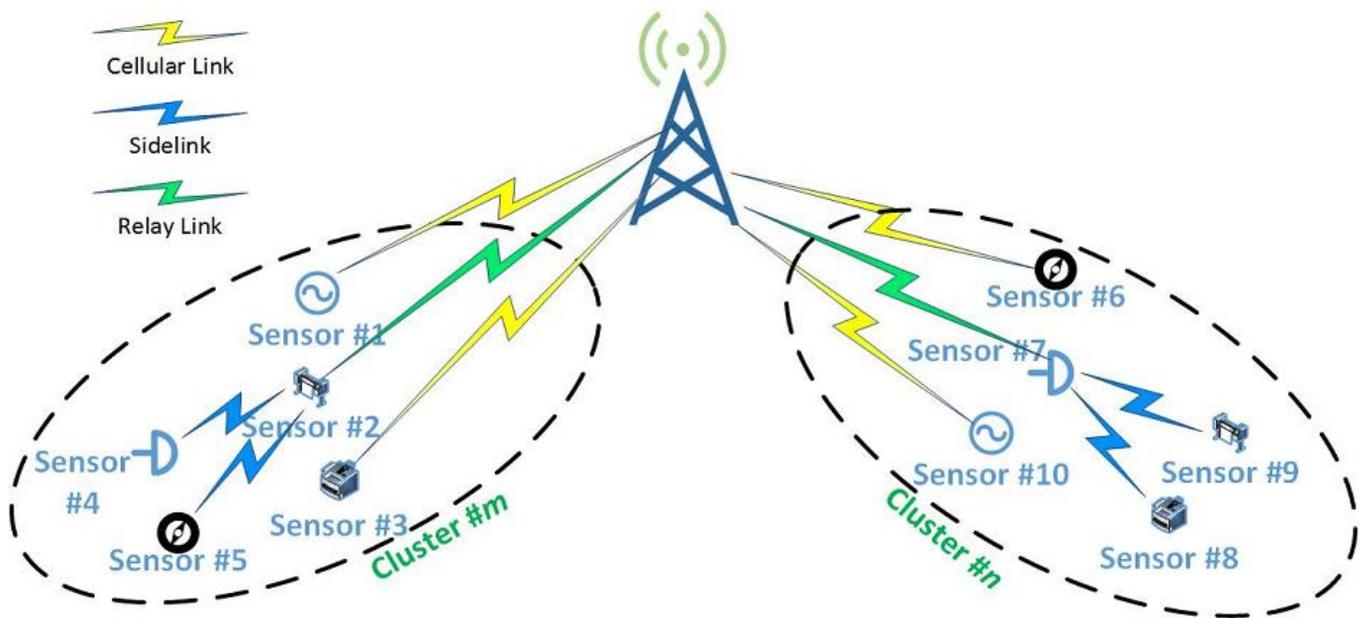

Figure 1: Our proposed approach allows the sensors with better propagation channel and enough battery level to relay uplink packets from other sensors.

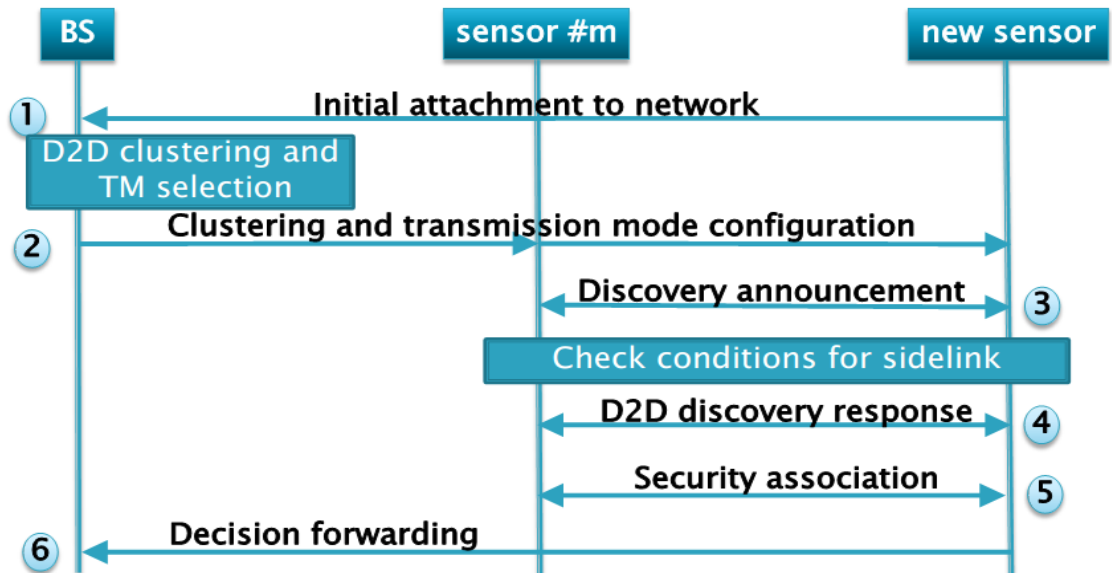

Figure 2: Initial attachment of a new sensor

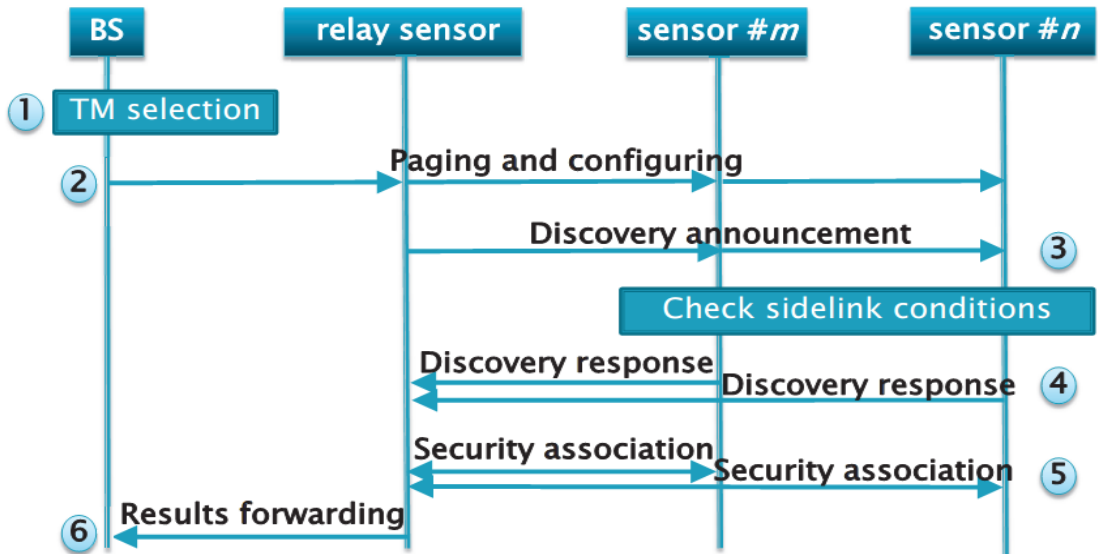

Figure 3: Procedure for TM update

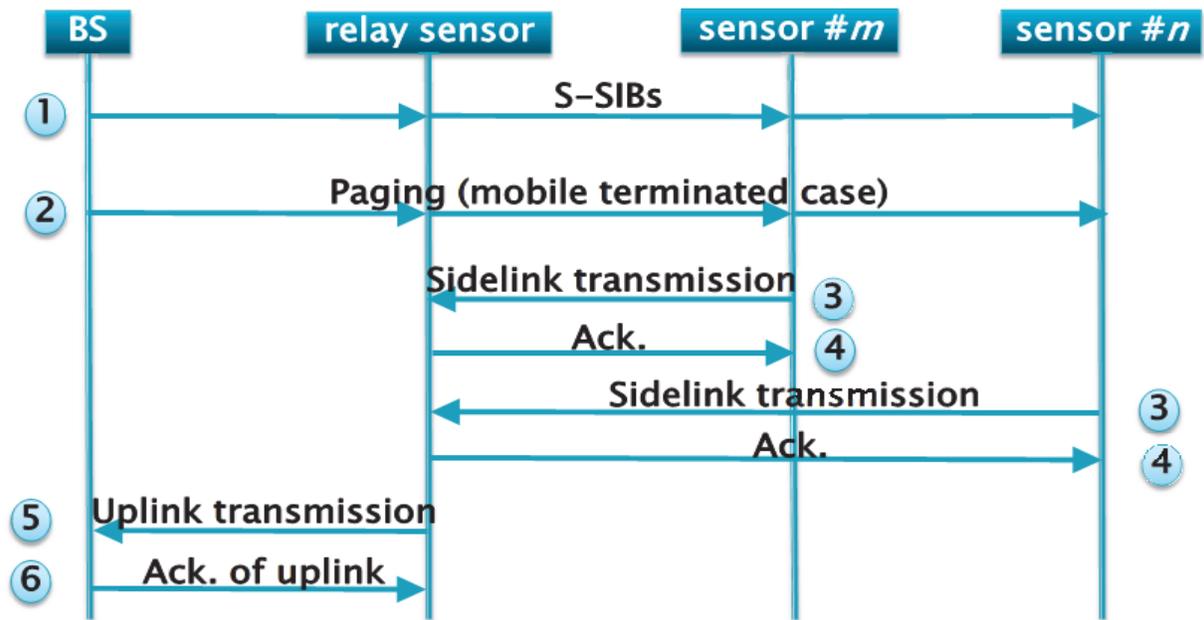

Figure 4: Uplink report procedure

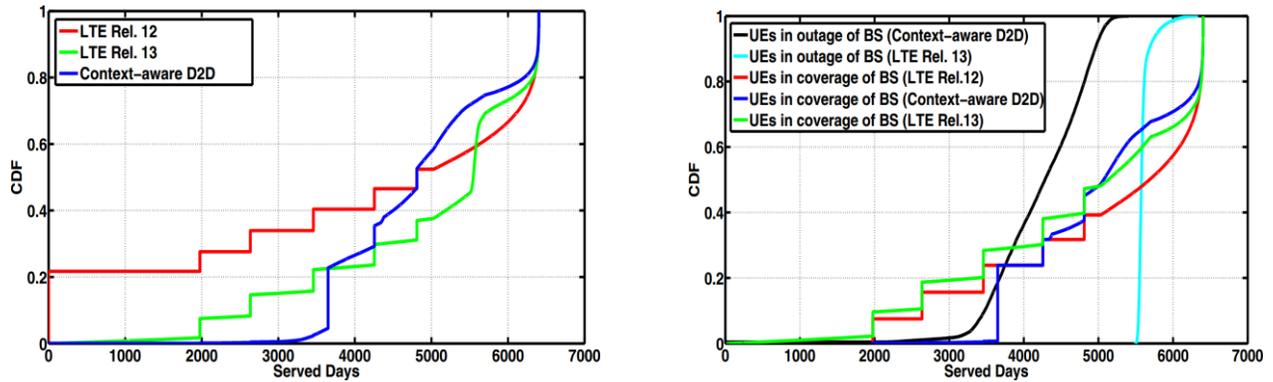

(a) CDF of served days of all deployed mMTC devices　(b) CDF of served days of mMTC devices from two different sets (in the coverage or outage of LTE)

Figure 5: System performance of mMTC devices w.r.t. their served days in uplink

TABLE I
DEVICE POWER CONSUMPTION PARAMETERS

| Parameter | Description | Value | Time duration if applicable |
|---|---|---|---|
| $P_{tx}$ | power consumption during transmission | 45% PA efficiency plus 60 mW/s for other circuitry | MCS and packet size related |
| $P_{rx}$ | power to receive packets from remote sensors | 100 mW/s | MCS and packet size related |
| $P_{paging}$ | power to receive paging command | 100 mW/s | 10 ms |
| $P_{clock}$ | power to obtain synchronization | 100 mW/s | 10 ms |
| $P_{cp}$ | power consumption during the control plane establishment procedure | 200 mW/s | 10 ms |
| $P_{sleep}$ | power consumption in sleeping mode | 0.01 mW/s | time of sensor staying in sleeping mode |
| $D_{rx}$ | number of DRX cycles per day | 4 times/day | |
| $C$ | battery capacity | 5 Wh | |
| $t$ | data reporting period | 150 s | |
| $I$ | packet size | 1000 bits | |
| $t_{DRX}$ | length of DRX cycle | 6 hours | |
| $T_{TMS}$ | periodicity of TM update | 1 day | |